# Anchoring-dependent bifurcation in nematic microflows within cylindrical capillaries

Paul Steffen,[1] Eric Stellamanns,[2] and Anupam Sengupta[1,a]

[1]*Physics of Living Matter, Department of Physics and Materials Science, University of Luxembourg, 1511 Luxembourg City, Grand Duchy of Luxembourg*

[2]*Deutsches Elektronen-Synchrotron DESY, Notkestraße 85, 22607 Hamburg, Germany*

[a] *Corresponding author:* anupam.sengupta@uni.lu

Capillary microflows of liquid crystal phases are central to material, biological and bio-inspired systems. Despite their fundamental and applied significance, a detailed understanding of the stationary behaviour of nematic liquid crystals (NLC-s) in cylindrical capillaries is still lacking. Here, using numerical simulations based on the continuum theory of Leslie, Ericksen and Parodi, we investigate stationary NLC flows within cylindrical capillaries possessing homeotropic (normal) and uniform planar anchoring conditions. By considering the material parameters of the flow-aligning NLC, 5CB, we report that instead of the expected, unique director field monotonically approaching the alignment angle over corresponding Ericksen numbers (dimensionless number capturing viscous v/s elastic effects), a second solution emerges below a threshold flow rate (or applied pressure gradient). We demonstrate that the onset of the second solution, a nematodynamic bifurcation yielding energetically degenerate director tilts at the threshold pressure gradient, can be controlled by the surface anchoring and the flow driving mechanism (pressure-driven or volume-driven). For homeotropic surface anchoring, this alternate director field orients against the alignment angle in the vicinity of the capillary center; while in the uniform planar case, the alternate director field extends throughout the capillary volume, leading to reduction of the flow speed with increasing pressure gradients. While the practical realization and utilization of such nematodynamic bifurcations still await systematic exploration, signatures of the emergent rheology have been reported previously within microfluidic environments, under both homeotropic (Sengupta et al., Phys. Rev. Lett. 110, 048303, 2013) and planar anchoring conditions (Sengupta, Int. J. Mol. Sci. 14, 22826, 2013).

**I. Introduction**

   As we continue pushing the boundaries of miniaturizing functional materials for both synthetic and biological applications, there is a growing need to better understand the dynamics and response of such materials at small scales (typically, micron and sub-micro scales). Liquid crystals (LCs), since their discovery in the late nineteenth century, have been at the forefront of scientific and technological breakthroughs over the last century, revolutionizing the way we see, sense, feel, touch and think [1-4]. From novel metamaterials and optofluidic based applications [5-8], to biological sensing and drug development [9-11], LCs are now an integral part of our daily lives, in more ways than we can perceive. A vast majority of these life-changing applications rely on the dynamic interaction of LCs – from molecular to microscopic scales – with their environments, mediated by the local pressure, boundary and confinement conditions. [12-14]. In addition to the confinement and surface anchoring, external fields play a key role in tuning the dynamic functionalities of LC materials. Hydrodynamic, electric and magnetic – the commonly studied external fields – have been central to the development of LC physics and applications, eliciting plethora of exotic dynamic attributes otherwise not observed in isotropic systems [16-23].

   Our ability to potentially harness these dynamical attributes relies on an unambiguous characterization of the microscale rheological properties of LC-s [13,14]. However, as demonstrated over the course of development of the field, such applied interests could often lead to fundamental insights, offering a rich landscape of unexpected dynamic manifestations, including generation [16] and sustenance [22,24] of transverse pressure gradients, anomalous viscoelasticity [25,26] and emergent biaxiality [27] and tunable flow shaping [28]. A key determinant of the effective rheological properties is the LC backflow mechanism, which only recently has been characterized in microfluidic setups [28,29], with potentially far-reaching ramifications in active and biological systems [30-32]. LC-s present distinct material advantages, thanks to the backflow-mediated coupling, over the isotropic counterparts. A coherent theory, with major contributions from Leslie, Ericksen and Parodi were achieved in the 1960-s. Despite almost immediate confirmation of the theory, the lack of computing power and experimental control had long hindered a more detailed understanding of even simplified stationary flow configurations. In the last two decades, these barriers have eased which enabled growing progress. LC microfluidics, in tandem with advanced computing tools, has allowed unprecedented insights through controlled flow experiments within channels spanning different surface, confinement and hydrodynamic constraints [13].

   Though the advent of LC microfluidics has propelled our understanding of the microscale LC flow behaviour, a bulk of the existing literature focuses on flows within channels with rectangular cross-sections [21, 28], revealing topologically distinct director profiles separated by transition flow speeds. However, a closer look at the natural systems would reveal that ubiquity of the cylindrical geometry, across a wide spectrum of biological and technological settings. Yet, studies on the flow within circular cross-sections (e.g., cylindrical capillaries) are relatively rare, thus necessitating the studies we have undertaken here. The significance of LC flows in cylindrical micro-capillaries, under natural and imposed constraints, have been demonstrated by the groups of Rey [33-36], Sluckin [37] and Yeomans [38,39], with a growing relevance of these studies in a range of applied and fundamental settings, both in the context of passive and active material systems [40-43]. Rey and co-workers have studied capillary Poiseuille flows using the Leslie-Ericksen-Parodi (LEP) theory for discotic nematic liquid crystals (NLC-s) [35]. Denniston et al. [38] and Batista et al. [40] performed Lattice-Boltzmann simulations allowing variations of the order parameter and obtained this transition as well, together with a shear thinning for the large flow velocity regime, while Ravnik et al. [39] have used lattice Boltzmann simulations of Landau-deGennes orientation tensor model to explore the behavior of active liquid crystals under Poiseuille flows. The LEP-based computations by Andersen et al. [41] confirmed experimental observations of different flow-induced director states [28], and explained the underlying energetics of the transitions between the director configurations. Zhou and Forest have studied capillary Poiseuille flows using the Doi-Marrucci-Greco model employing a second-moment tensor description of the orientation distribution [44].

A unified description of nematodynamics, bridging the existing gap between the classical LEP theory and the Doi theory, was first proposed by Tsuji and Rey [33]. By accounting for the relative contribution of the short- and long-range elasticities in shear flows, quantified as their ratio, $R$ (Reactive parameter), the authors presented a rheological phase diagram with four distinct regimes under fixed planar director boundary conditions, spanning Deborah ($De$) and Ericksen ($Er$) numbers. Of the four regimes, the elastic-driven steady state corresponds to the LEP solution for the non-aligning nematics, while the viscous-driven steady state corresponds to the flow aligning behaviour at higher $Er$ numbers. Since the two dimensionless numbers respectively capture the ratio of the viscous to short range elastic effect ($De$) and the ratio of the viscous effect to long-range elastic effect ($Er$), their mutual ratio results to the Reactive parameter, $R = Er/De.$ The asymptotic limits of the in-plane shear flows yield the LEP theory (for $R \to \infty$) and the Doi theory (for $Er \to \infty$). Employing a Landau-de Gennes model, Alonso et al. studied Couette flow of NLC-s using asymptotic methods and numerical bifurcation theory [37]. In addition to confirming the four regimes, their results highlighted the existence of a Takens-Bogdanov point, a high-order singularity at the coincidence of a Hopf bifurcation and a limit point. It may be worthwhile to note that the applied technological interests (in particular, the processing of high performance fibers and films) which have guided investigations on the dynamical rheology of LC-s, have mostly focused on LC polymer materials under Poiseuille flows. However, complementary investigations on the flow of non-polymeric LC materials within cylindrical capillaries are still underexplored. The lack of systematic micro-scale studies of LC flows within cylindrical capillaries, or of the associated role of surface anchoring conditions (how LC molecules orient on the confining boundaries), offer opportunities to gain fundamental insights on dynamical states and nematodynamic bifurcations, with potential ramifications on the flow-structure coupling in a range of soft, active and biological systems.

In this paper we investigate stationary flows of NLC-s within cylindrical capillaries possessing strong homeotropic (normal) and uniform planar anchoring conditions using numerical simulations based on the continuum theory of Leslie, Ericksen and Parodi (LEP theory). By spanning flow speeds over two orders of magnitude, we identify that below a threshold flow rate, a second flow-director solution emerges, in addition to the expected director field orientation. By systematically varying the Ericksen number (Eq. 1) of our capillary flows *in silico*, we report experimentally tractable nematodynamic bifurcation points for nematic 5CB, as alternate solutions in the apparent viscosity emerge at the threshold pressure gradient. Although our simulations are carried out for a single capillary diameter (2*r* = 831 μm) to reflect the dimensions in our ongoing experiments, the results presented here hold good for micron-scale cylindrical capillaries, with appropriate scaling of the relevant dimensionless numbers. We found that the bifurcation coordinates on a pressure-viscosity parameter space depend on the nature of surface anchoring on the walls of the cylindrical capillary, and on the driving mechanism of the flow. For strong homeotropic anchoring conditions, the flow bifurcation occurs at a pressure gradient of 14,7 Pa/m (flow rate 3.5 nl/s) and apparent viscosity 50 mPas. In contrast, for the capillaries with planar boundary conditions, the bifurcation point is determined, in addition, by the driving mechanism of the flow, i.e. pressure-driven flows (constant pressure) v/s volume-driven flows (constant volume rate). For the pressure-driven flows, the bifurcation occurs at 5.9 Pa/m (corresponding flow rate is 2.2 nl/s) and apparent viscosity 31 mPas; whereas for volume-driven flows, the bifurcation takes place at a flow rate of 1.8 nl/s (corresponding pressure gradient is 8 Pa/m) and an apparent viscosity of 51 mPas. Overall, the onset of the anchoring-dependent apparent viscosity of nematic 5CB at a given pressure gradient (or flow rate) is coupled to the alternate local director orientations with similar energetic costs. For homeotropic surface anchoring, the second director field orients against the alignment angle close to the capillary center; while in the uniform planar case, the alternate director field extends throughout the capillary volume. Although experiments to systematically capture the simulated bifurcations are underway, signatures of the same could be found in author's previous works [28,45].

## II. Theoretical Background

Based on the Frank theory of LC-s [46], Ericksen, Leslie and Parody proposed a continuum theory for nematic flows, taking viscosity as well as the interaction between flow and director field into account. In the following, we will refer to it as the LEP [47-49]. It´s dominant parameter is the Ericksen number, *Er*:

$$Er = \frac{\eta \cdot v \cdot L}{k} \qquad (1)$$

where $\eta$, $v$, $L$ and $k$ are the characteristic dynamic viscosity, flow speed, length scale and elasticity. For small *Er* numbers, the elastic interactions dominate, for large numbers the viscous ones. A predicted shear alignment angle of the director field in the limit of high *Er* numbers – the Leslie angle - was confirmed through experiments in Couette flow using polarization optics by Gähwiller [50]. The Leslie angle depends on the material parameters as described in Section III. For stationary flows in cylindrical geometry, Atkin provided the ordinary differential equations which were used by Tseng et al. to calculate numerical solutions for the p-azoxyanisole and experimentally tested by Fishers and Fredrickson [12,51,52]. For homeotropic anchoring, the experimentally measured viscosity decreased monotonically with increasing Ericksen numbers. It scaled perfectly with *Er* for diameters between 78 µm and 516 µm and 20 < *Er* < 100,000. Using stationary Couette and channel flows, Pieranski and Guyon had demonstrated existence of multiple director configurations due to bifurcation under uniform planar anchoring with surface orientation perpendicular to the flow direction [53-55]. While Couette flows gave a quantitative agreement between the LEP and the experiments, for channel flows qualitative agreement with the LEP was concluded LEP. By employing non aligning nematics, oscillating flow and electric fields magnetic fields, several further instabilities and bifurcations were obtained [56-59]. Despite the simplifications made by Lima and Rey to numerically capture the stationary states of flowing discotic nematics within a cylindrical geometry [35], a plethora of nonlinear phenomena could be captured. Shear thinning, shear thickening, distinct viscosity maxima and minima as well as multiple stable solutions with mutually different slope signs were obtained.

The present investigation follows up on the work of Lima and Rey [35], to specifically capture the flow behaviour of rod-shaped NLC-s flowing within a cylindrical capillary, the role of surface anchoring on such microsscale flows, and the possible dependence of the flow states on the driving mechanisms of the flow (pressure-driven v/s volume-driven). Based on our understanding of the existing gaps, we have chosen study rod-shaped NLC-s, which possess fewer degrees of freedom than what has been reported for the discotic nematics. Through an in-depth numerical treatment of this basic system, the results of this work will provide a concrete perspective on designing and optimizing more complex cases and potential applications. With this in mind, we tackle the NLC flow problem along the lines of the work by Lima and Rey, however for rod shaped NLC-s (instead of discotic ones) flowing under different anchoring conditions.

**III. Methods and Materials**

The flow geometry in our work is sketched in Fig. 1. We capture the flow dynamics by solving two ordinary differential equations obtained by Atkin [51], obtained directly from the LEP under the assumptions described in Section III A. The capillary diameter, flow rates and the material properties of our system are chosen to reflect those in our experiments, which will be reported in a forthcoming paper. The paper is organized in the following manner. In the Materials and Methods (Section III), we describe the equations governing our numerical simulations (Section III a), the material parameters applied (Section III B), and finally our numerical treatment (Section III C). The Results section is divided into Section IV A for strong homeotropic anchoring and Section IV B for strong uniform planar anchoring oriented along the flow direction. For both anchoring types, we present the director fields, the shear rates and the apparent viscosities at Ericksen numbers spanning the elastic and the viscous regimes. In Section V A, we discuss the implications and plausibility of the solutions obtained, and Section V B suggests existing experimental evidence of our solutions. Finally, we conclude the paper in Section VI and provide a short perspective on potential applications of our findings.

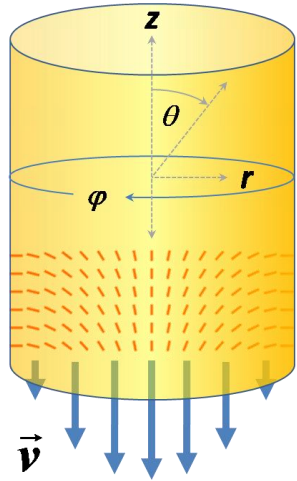

**FIG. 1. Flow geometry and anchoring.** The investigated geometry is a circular micro-capillary, shown here with the cylindrical coordinate system. The symmetry axis of the capillary is denoted by $z$, the radial distance from this axis by $r$, the tilt angle relative to the symmetry axis by $\theta$, and the azimuthal angle by $\varphi$. The red line-segments shown in the lower half of the capillary indicate the orientation of the rod shaped 5CB molecules, and the blue arrows at the bottom present the flow direction.

**A. Equations to solve**

The geometry and the relevant coordinate system are described in Fig. 1. Following Atkin [51], we apply the following conditions on the flow velocity $\vec{v}$ and the director field $\vec{n}$ [8]:

$$v_r = 0 \qquad v_\varphi = 0 \qquad v_z = v(r) \qquad (2)$$

$$n_r = \sin(\theta(r)) \qquad n_\varphi = 0 \qquad n_z = \cos(\theta(r)) \qquad (3)$$

where $\theta(r)$ is the director tilt with respect to the symmetry axis $z$ of the capillary. This ansatz satisfies the continuity equation, i.e. it implies incompressible flow, and imposes rotational symmetry for $\vec{n}$ and $\vec{v}$,. Leading us to:

$$g(\theta) \cdot v'(r) = -\frac{1}{2} \cdot a \cdot r + \frac{b}{r^2} \tag{4}$$

$$2 \cdot f(\theta) \cdot [\theta''(r)]^2 + \frac{2}{r} \cdot f(\theta) \cdot \theta'(r) - \frac{k_{11}}{r^2} \cdot \sin[2 \cdot \theta(r)]$$

$$-[\cos(2 \cdot \theta(r)) - \cos(2 \cdot \theta_0)] \cdot \lambda_3 \cdot v'(r) = 0 \tag{5}$$

with,

- $2 \cdot g(\theta) := 2 \cdot \mu_1 \cdot \sin^2(\theta) \cdot \cos^2(\theta) + (\mu_5 - \mu_2) \cdot \sin^2(\theta) + (\mu_6 + \mu_3) \cdot \cos^2(\theta) + \mu_4$ (6)
- $\mu_i :=$ Leslie coefficients for the viscosity ($i$ = 1 - 6)
- $v'(r) := \frac{\partial}{\partial r}[v(r)]$ is the shear rate and $v(r)$ the low velocity (7)
- $a$ is a constant; due to Atkin [51] it denotes the axial pressure gradient *dp/dz*, where *p* denotes the pressure applied across the capillary
- $b$ is a constant as well; Atkins used it in the case of the flow between two coaxial cylinders; in our case, where the inner cylinder not exists, holds $b = 0$. Else, the stress and pressure values would diverge to infinity;
- $f(\theta) := k_{11} \cdot \cos^2(\theta) + k_{33}\sin^2(\theta)$ (8)
- $k_{11} :=$ elastic coefficient for splay; $k_{33} :=$ elastic coefficient for bending
- $\theta'(r) := \frac{\partial}{\partial r}[\theta(r)]$ ; $\theta''(r) := \frac{\partial^2}{\partial r^2}[\theta(r)]$ (9)
- $f'(r) := \frac{\partial}{\partial r}[f(r)]$ (10)
- $\cos(2 \cdot \theta_0) := -\frac{\lambda_1}{\lambda_2}$ (11)

  the aligning angle or Leslie angle, i.e. the limiting angle at large shear rates where elastic terms are negligible, and are dominated by the viscous terms;
- $\lambda_1 := \mu_2 - \mu_3 :=$ rotational viscosity (12)
- $\lambda_2 := \mu_5 - \mu_6 :=$ torsional coefficient (13)

Lima and Rey [35] gave an equivalent, regrouped and dimensionless version of eq. (4), (5). In the static case, this is characterized by $a = v'(r) = 0$, for which an analytical solution could be obtained for eq. (5) [60,61].

**B. Material Parameters**

The material parameters used here were obtained from Sengupta et al. and Dhar et al. [24,62]. The elasticities of the director field $\vec{n}$ for splay $k_{11}$ and bend $k_{33}$ are taken as 6.4 pN and 9 pN respectively. The coefficients for twist and for saddle splay $k_{22}$, $k_{24}$ do not show up in the final equations. For the Leslie coefficients $\mu_i$ we assumed in units of mPas : $\mu_1 = -6; \mu_2 = -81.2; \mu_3 = -3.6; \mu_4 = 65.2; \mu_5 = 64; \mu_6 = -20.8$. If the resulting $\lambda_{1,2}$ satisfies the condition:

$$1 < \left|\frac{\lambda_1}{\lambda_2}\right|, \tag{14}$$

the alignment angle described above cannot be real and the stationary solution ansatz followed here works only for small Ericksen numbers (with no stationary solution at high *Er*), signifying the tumbling NLC-s. The material parameter for 5CB lead to an aligning behavior with a Leslie angle of 11.89° [24].

The term $g(\theta)$ in eq. (4) is analogous to the viscosity in the corresponding equation in the Hagen-Poiseuille flow, except it is dependent on the angle $\theta(r)$ and therefore on $r$ as well, it is called anisotropic viscosity. It's dependence on $\theta$, i.e. the tilt angle of the director, is plotted in Fig. 2. The value for $\theta = 0°$ (director in the shear plane and perpendicular to the velocity gradient) is known as the Miesowicz viscosity $\eta_1 = \frac{1}{2} \cdot (\mu_3 + \mu_4 + \mu_6)$, with a value of 20.4 mPas for our chosen material. The value for $\theta = 90°$ (director in the shear plane and parallel to the velocity gradient), corresponds to the Miesowicz viscosity, $\eta_2 = \frac{1}{2} \cdot (\mu_4 + \mu_5 - \mu_2)$, with a value of 105.2 mPas for our material set [24]. The remaining Miesowicz viscosity (director perpendicular to the shear plane) can be computed as $\eta_3 = \frac{1}{2} \cdot \mu_4$, with a value of 32.6 mPas for our chosen material. The corresponding shear is however prohibited by the symmetry assumptions eq. (2) and eq. (3).

The coefficients $k_{ii}, \eta_i$ represent the anisotropy of the NLC, and thus should be appropriately accounted for in the computing the Ericksen number in Eq. (1). Because the elastic coefficients $k_{11}$, $k_{33}$ are both always present in our configuration and $\mu_4$ represents the isotropic Leslie coefficent, with the corresponding viscosity comparably less dependent on the temperature, we choose $k := k_{11} + k_{33}$ as the characteristic elasticity and $\eta_3 = \frac{\mu_4}{2}$ as the characteristic viscosity in eq. (1). Finally, we choose our main capillary diameter (831μm) as the characteristic length scale and the average velocity as the characteristic one, which cause the Ericksen numbers of the simulations to be 3.2 times higher than the simulated flow rates in units of [nl/s].

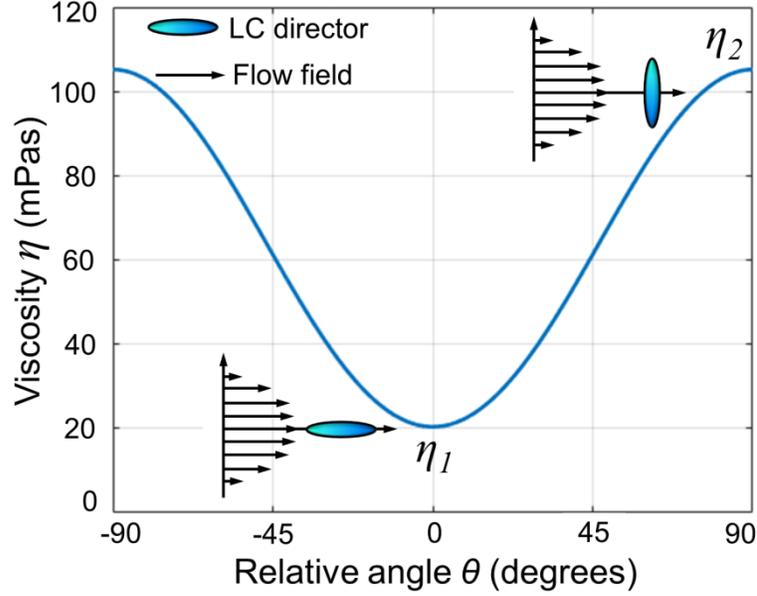

**FIG. 2. Miesowicz viscosities.** Due to the anisotropy of the viscosity coefficients, the apparent viscosity of the flowing NLC depends on the flow-induced director tilt angle. It is determined by the function $g(\theta)$ in eq. (4). The Miesowicz viscosities $\eta_1$ and $\eta_2$ are described within the text [15,24].

### C. Numerical approach

Following the approach by Tseng et al. [52], we solve Eqs. (4) and (5) by imposing $\theta(r = 0) = 0°$, i.e. $\vec{n}$ to be parallel to the $z$-axis in the capillary center. We considered two anchoring cases: in section IV a, we investigate $\theta(r = R) = 90°$, i.e. the homeotropic anchoring, and in section IV b $\theta(r = R) = 0°$, i.e. uniform planar anchoring with surface anchoring parallel to the flow direction. Since $\theta(r = R)$ is not influenced by the slope of $\theta(r)$ in the neighborhood, strong anchoring holds in both cases. For obtaining the director field at given values of capillary radius $R$, $\theta(r = 0)$, $\theta(r = R)$, the flow rate $\dot{V}$, the Leslie coefficients $\mu_{i=1-6}$ and the elasticities $k_{11}$, $k_{33}$, we first eliminate $v'(r)$ from eq. (5) by eq. (4). Then we impose a test value for $a$, i.e. the axial pressure gradient $dp/dz$, into eq. (5) and integrate it, which gives us a boundary problem that results from imposing $\theta(r = 0)$ and $\theta(r = R)$. By adapting the boundary problem into an initial value problem using the shooting method [63], through imposing $\theta$ and testing with $\theta'$ either at $r = 0$ or at $r = R$ until the desired adjacent angle is obtained. The first alternative was used to obtain solutions where the sign of $\theta$ changes within the integration range. Due to the properties of eq. (5), we could not apply exactly $r = 0$, therefore we begun the integration at $0 < r \ll R$. The corresponding flow rate $\dot{V}$ is then obtained from $\theta(r)$ by using equation (4) and applying the no slip condition $v(r = R) = 0$ at the surface of the capillary:

$$\dot{V} = 2\pi \cdot \int_0^R r \cdot \left\{ \int_R^0 \frac{a \cdot x}{2 \cdot g(\theta(x))} dx \right\} dr \qquad (15)$$

We iterate this procedure with different values of $a$ until the desired flow rate $\dot{V}$ is obtained, independently using software packages Mathematica and Matlab which gave identical results. Lima and Rey [35] had undertaken a dimensionless equivalent of our eq. (4), (5) using finite element method, with different material parameters corresponding to those of the discotic nematics. They extended the calculations to higher Ericksen numbers to arrive at multiple rotations of the director field confined between between the capillary center and the walls. We have validated our numerics by selectively repeating the calculations of Lima and Rey, plugging in corresponding material parameters into our numerics which yielded equivalent results.

## IV. Results

For an isotropic fluid one would expect a parabolic flow profile for the Hagen-Poiseuille flows with the shear rate increasing strictly linear with the radial distance, $r$. A nematic phase pertubs this profile due to the uniaxial anisotropy. Firstly, its effective viscosity at a given location depends on the angle $\theta$ between the director field and the flow direction. The anisotropic viscosity $\eta(\theta)$ is plotted in Fig. 2 for the material parameters chosen here. Secondly, owing to the flow-director coupling, the flow field distorts the director field and at low Ericksen numbers, nematic backflow kicks in whereby the director field impacts the flow field. In the following sections we will describe the resulting director fields, shear rates and the emergent dynamic viscosity over different flow rates (Ericksen numbers) under strong homeotropic as well as strong uniform planar anchoring.

### A. Homeotropic anchoring

Fig. 3 presents the spatial variation of the flow-induced director field within the cylindrical capillary possessing strong homeotropic anchoring, i.e. the director at the surface is oriented strictly perpendicular to the surface. The resulting director tilt angles $\theta(r)$ are plotted for different flow rates (in units of nl/s, i.e. nanoliters per second), as indicated by the different legend hues. The corresponding director field projection for each case is presented by the panels in Fig. 4. The red line-segments in Fig. 4 indicate the variation of the director tilt angles, while the solid black line between the red line-segments quantify the local shear rates (the ticks and accompanying numbers indicate the shear rates in $1/s$). The two thick vertical black flanking either sides denote the capillary wall in each panel. As shown, the shear rate in the capillary center is always zero, which follows from eq. (2) for a differentiable velocity profile. In the static case (flow speed of 0.0 nl/s), the shear rate is zero everywhere, as depicted in upper row of Fig. 4 (middle panel). Under homeotropic boundary conditions, two different non-singular static director configurations can spontaneoulsy emerge within our capillaries (singular solutions are ruled out for capillary dimensions considered here): the "escape to the top" and the "escape to the bottom" configurations. The results presented in Fig. 3 and Fig. 4 focus exclusively on the bottom escape configuration. Positive (negative) flow rates denote the direction of the NLC microflow parallel (antiparallel) with respect to the escape direction.

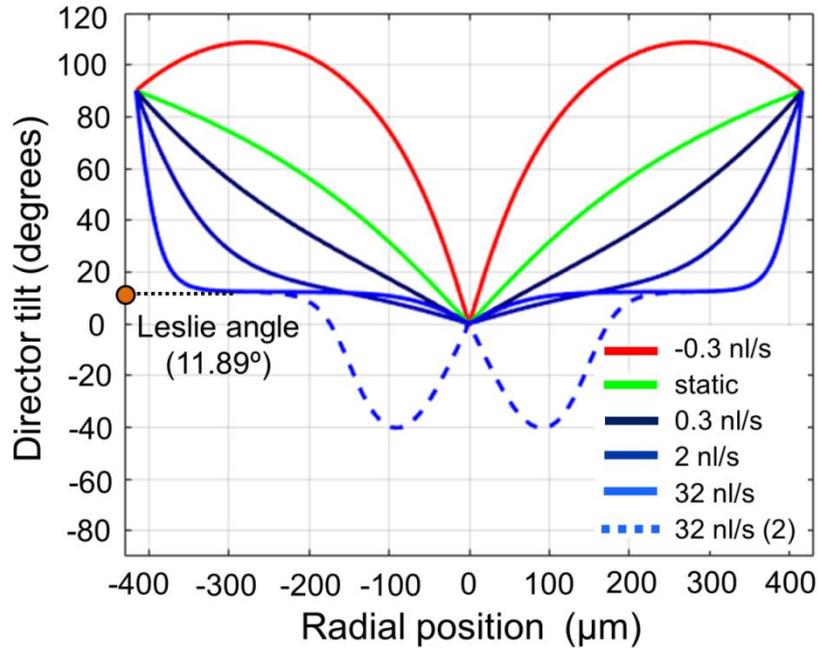

**FIG. 3. Flow-induced director tilt under strong homeotropic anchoring.** The spatial variation of the numerical director tilt angles $\theta(r)$, for strong homeotropic anchoring, are plotted for different flow rates shown with different hues (in units of nl/s, i.e. nanoliters/second). The corresponding Ericksen numbers *Er* are 3.2 times the flow rate in nl/s.

The escape director field for the nematics at rest (green line in Fig. 3) is solely shaped from the competition of elastic forces, namely splay and bend, excluding twist due to the symmetry inposed by eq. (2) and (3). The shear involved in a moderate flow rate of +/- 0.3 nl/s already bends the director field, however the effect depends on the direction of the imposed flow relative to the escape direction. For positive flow rates, i.e., the flow is parallel to the direction of escape, the director configuration is pushed to lower angles (visualized in Fig 4, upper row right panel), whereas for negative flow rates the director tilt reaches higher angles and a non monotonic depndence on $r$ is captured (visualized in Fig 4, upper row left panel). The negative flow rate induces a stronger director deformation than the positive one and is even able to orient the director field against the initial escape direction in the outer part of the capillary. The absolute value of its shear rate shows a strong slope in the immediate vicinity of the capillary center, i.e., at small $r$, reflecting strong local gradient of elastic distortions for a small director angle $\theta$. Our simulations reveal that the high local shear rate starts to flatten at moderate $r$, since $\theta(r)$ increases quickly, and thereby the anisotropic viscosity as well. At moderate positive flow rate (+0.3 nl/s), the director angle $\theta$ increases gradually which causes the flattening of the shear rate over larger distances from the cpaillary center. From the surface towards the center, the director tilt however varies more drammatically, resulting in stronger gradients in director field. This leads to local anisotropic viscosities which stabilize the maximum shear rate away from the capillary wall, between the center line and the wall (Fig 4, upper row right panel). The positve flow rates tilt angles $\theta(r)$ are, on average, further away from 90° which explains its lower viscosity compared to the negative flow rate shown in Fig. 9a.

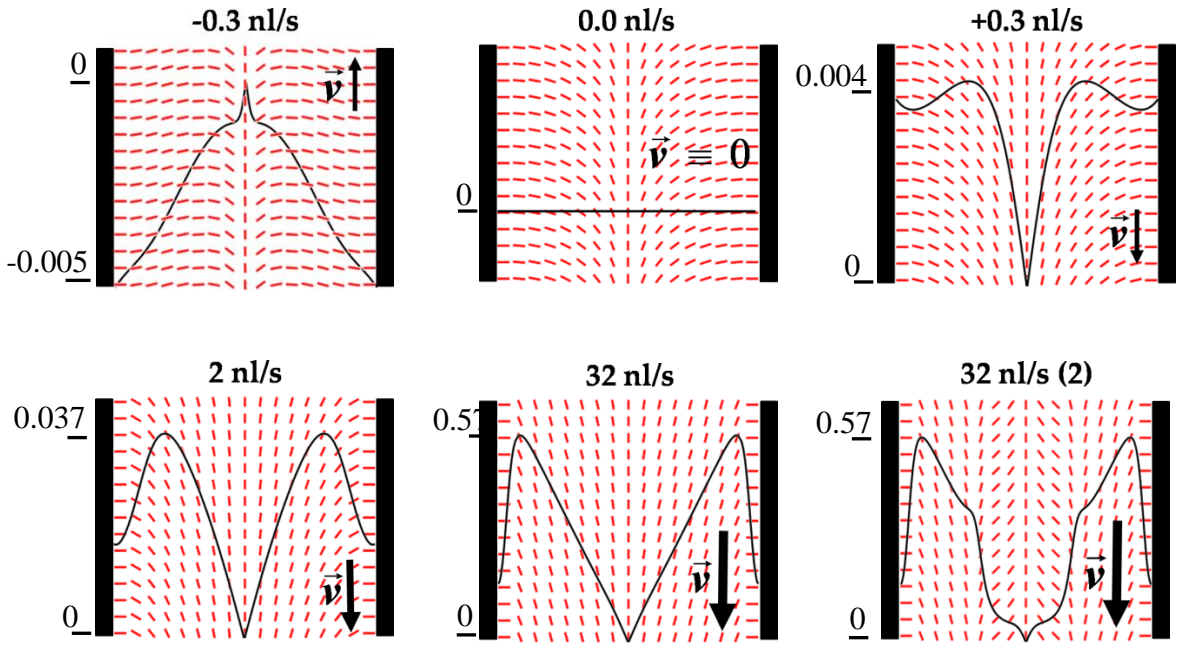

**FIG. 4. Flow-induced director field and emergent shear rates (homeotropic anchoring).** The red line-segments capture the flow-induced director tilt angle, corresponding to the conditions plotted at Fig. 3. The black line quantifies the emergent shear rates, which remains zero at the capillary center. The plots are normalized for having an equal maximum height, with the exception of the static case.

Upon increasing the flow rate to $\dot{V}$ = 2 nl/s, the director tilts towards the flow direction, i.e. the tilt angles $\theta(r)$ decrease everywhere. This effect is stronger close to the confinement surfaces than at the center, where the shear rates are smaller and consequently, the shear-induced torques as well. In combination with the strong homeotropic anchoring condition, this results in stronger tilt angle gradients close to the surfaces, ultimately resulting in a sharper shear rate maximum, and relatively larger shear rate decay toward the surfaces (Fig 4, lower row left panel). The apparent viscosity drops due to the smaller tilt angles as captured in Fig. 9a, in agreement with the commonly observed shear thinning behaviour [45].

For $\dot{V}$ = 32 nl/s, we found two different solutions. The first one, denoted by the bold line in Fig. 3 (light blue hue) shows a steady title angle of of around $\theta_0$ =11.89°, which does not vary further. $\theta_0$ is called the equilibrium Leslie angle, which can computed also from the Leslie coefficents $\alpha_i$ as described in Section III. For $\theta = \theta_0$, the elastic and viscous torques balance out each other. Therefore $\theta_0$ is the limiting tilt angle at high Ericksen numbers, wherein the contribution of the elastic torque on the director configuration is negligible [34,50]. The second solution for $\dot{V}$ = 32 nl/s, visualized by the dashed blue line, exhibits a non-monotonic dependence of the tilt angle on the local position within the capillary. The tilt angle dips to -40°, in agreement with previous reports [52]. The smallest flow rate which triggers the onset of this second solution is the point of nematodynamic bifurcation, which we found $\dot{V}$ = 3.5 nl/s. The apparent viscosity corresponding to the two numerical solutions (shown in Fig. 9a), as well as the corresponding director fields, can be clearly distinguished at the bifurcation. At low flow rates, the two solutions do not meet there each other, in close agreement with the phenomenon reported by Lima and Rey for the stationary flow of discotic nematics within circular capillaries [35]. For higher flow rates, the viscosities of the two solutions merge quickly together: at 5 nl/s, the two solutions can hardly be distinguished. We obtained the second solution up to our highest considered flow rates around $\dot{V}$ = 150 nl/s. As visible in Figs. 3, 4, the two director solutions (for 32 nl/s) are similar to each other in the outer parts of the capillary (large values of *r*), however vary considerably close to the capillary center line. Consequently, the shear rates mirror the director field, with emergence of strong gradients in the shear rate at the capillary center line.

**B. Uniform planar anchoring**

Figure 5 depicts our simulation results for the NLC flows within cylindrical capillaries possessing strong uniform planar anchoring, analogous to the results on the homeotropic case (presented in Fig. 3). The surface orientation of the nematic director is parallel to the flow direction. For this combination of geometry and anchoring condition, there is no spontaneous symmetry breaking at zero flow rates, thus we do not differentiate between positive or negative flow rates. The solutions presented in Fig. 5 are obtained under the condition:

$$\frac{\partial}{\partial r}\big(\theta(r)\big) \geq 0 \tag{16}$$

around the capillary center. Spatially, the director tilt spans the surface-anchored orientation, 0°, to the equilibrium Leslie angle, ~12°, thus covering a much smaller angular range relative to the homeotropic case. Overall, the spatial variation of the director tilt follows the same trend across the different flow rates, which are represented by differently hued plotted lines in Fig. 5. Complementing the spatial director tilt, Fig. 6 visualizes the corresponding shear rates for the static and dynamic cases (left and right panels respectively). At zero flow, the director field is completely uniform, hence the shear rate = 0. The higher the flow rates, the higher the tilt angles within the capillary, which gradually approach the equilibrium Leslie angle as the flow rate is enhanced. At the highest flow rate shown here (128 nl/s), the direct field is spatially aligned to the Leslie angle (red line in Fig. 5) almost through the entire capillary, except at the neighborhood of $r = 0$ (along the center line) and at $r = R$ (at the capillary wall) where the tilt angle is imposed to be zero (as discussed in Section IIIC). At the boundary and at the center, the angle drops with a certain slope down to zero, at the center with a smaller slope, due to the lower shear rates there. Similar director fields were reported for the Hagen-Poiseuille flow of discotic nematics under planar anchoring by Lima and Rey [35].

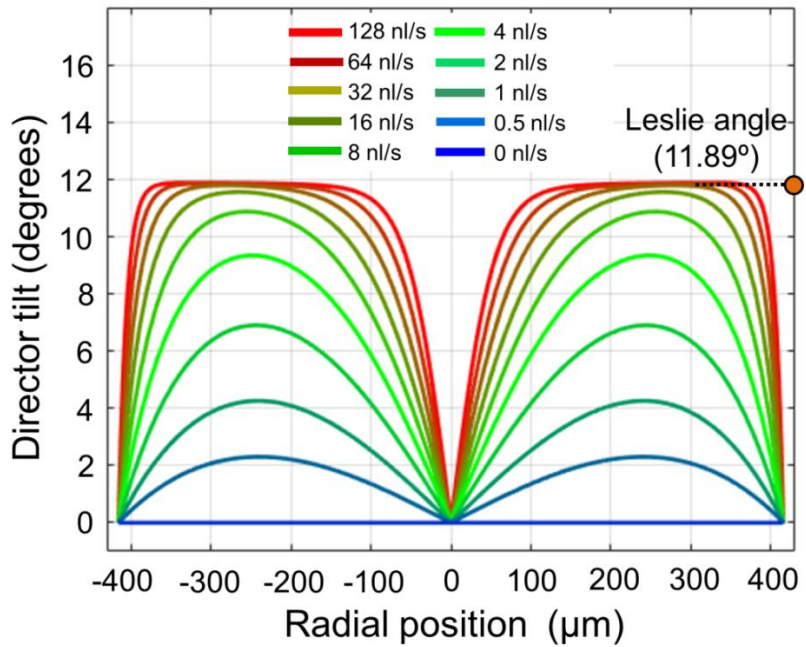

**FIG. 5. Flow-induced director tilt under strong planar anchoring (positive tilt).** Analogous to Fig. 3, the spatial variation of the flo-induced director tilt is plotted over different flow rates. The maximum tilt angle in uniform planar anchoring is the Leslie angle, ~12°, which is much smaller than the maximum flow-induced tilt angle under homeotropic anchoring conditions (110°, Fig 3).

Owing to the pronounced role of backflow close to the capillary walls, this director configuration is reflected characteristically in the shear rate profile (Fig. 6, right panel). The simulated shear rate profile shows qualitative similarity with that of an isotropic fluid underdoing Hagen-Poisseuille flow, however two specific upward kinks appear here due to the flow-director coupling. The first kink is observed prominently close to the boundary (due to backflow), and a second minor upward kink is captured at the center due to the minute enhancement of local flow rate at the center (title ~0°) relative to the vicinity where the title angle is ~ 12°. The apparent viscosity and pressure gradients emerging within the planar flow configurations are plotted in green hue in Fig. 9a,b ( denoted by the legend *planar anchoring, positive tilt angles*). Due to the increase of the tilt angles with increasing flow rates, the viscosity in this case shows a small positive slope. The slope is comparatively large at small flow rates and shrinks for larger flow rates, finally saturating to constant viscosity as the director tilt reaches the Leslie angle.

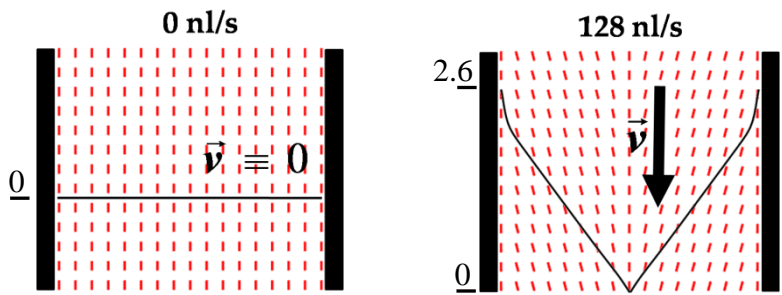

**FIG. 6. Flow-induced director field and emergent shear rates (planar anchoring).** For two of the flow rates in Fig. 5, static case (left panel) and 128 nl/s (right panel), the director tilt orientation is visualized by the red line-segments. The shear rates are denoted by the black lines overlaying the director tilt. The static case shows a completely homogenous director field, with zero shear rate. For the dynamic case (128 nl/s), the flow-induced change in director title is rather small (from 0 to 12°), however a pronounced kink in the shear rate could be seen close to the wall due to the local backflow effect enhanced by the no-slip boundary conditions.

In Fig. 7, we present the computed director and emergent shear rate profiles for the complementary uniform planar case, as given by:

$$\frac{\partial}{\partial r}\big(\theta(r)\big) \leq 0 \qquad (17)$$

around the capillary center line. For the smallest flow rate (1.83 nl/s), we obtain a steep negative gradient of the tilt angle around the capillary center line, accompanied by a more gradual negative tilt angle, with a minimum tilt at around -120° located at a distance of 200 μm from the center. Due to the anchoring condition, the tilt angle increases steeply after this minimum, ultimately reaching 0° at the center line. The director configuration suggests strong elastic deformations, which couple with the flow field to lead to a complex shear rate profile shown in Fig. 8 (upper left panel). Upon increasing the flow rate to 1.95 nl/s, we capture a similar trend, however with considerably weaker slopes (Figs 7 and 8, upper right panel). On average, the tilt angles for the larger flow rates deviate further from 90°, i.e. the anisotropic coupling between the flow and director field weaken at higher flow rates. Thus, one expects a lower apparent viscosity for a flow rate of 1.95 nl/s (37 mPas) than at 1.83 nl/s (51 mPas), as demonstrated in Fig. 7 by the dashed green and blue plots. Interestingly, when we look at the resulting pressure gradient, we can conclude that higher the flow rate, lower is the pressure gradient (8 Pa/m for lower flow rate v/s 6.2 Pa/m for the higher flow rate). Such stationary flow characteristics have, to the best of our knowledge, been reported for the first time here. A nematodynamic regime in which the pressure drops when the flow rate increases can reach stationary flow state only if the flow is externally imposed. If external pressure is imposed (pressure-driven flows), the system turns unstable, as presented in the apparent viscosity v/s pressure gradient plot in Fig. 9a (red hue, planar anchoring with negative tilt angle). The onset of this solution type is above zero, at ~1.8 nl/s (Fig. 9b), resulting in a bifurcation for the uniform planar anchoring case with its branches having different viscosities, i.e. the two branches do not meet each other. The unstable part, where the pressure gradient drops with increasing flow rates, is shown with dotted lines. For higher flow rate, the viscosity drops further, with its slope decreasing continuously until it meets the saturated viscosity which essentially corresponds to the expected first solution shown in Fig. 5.

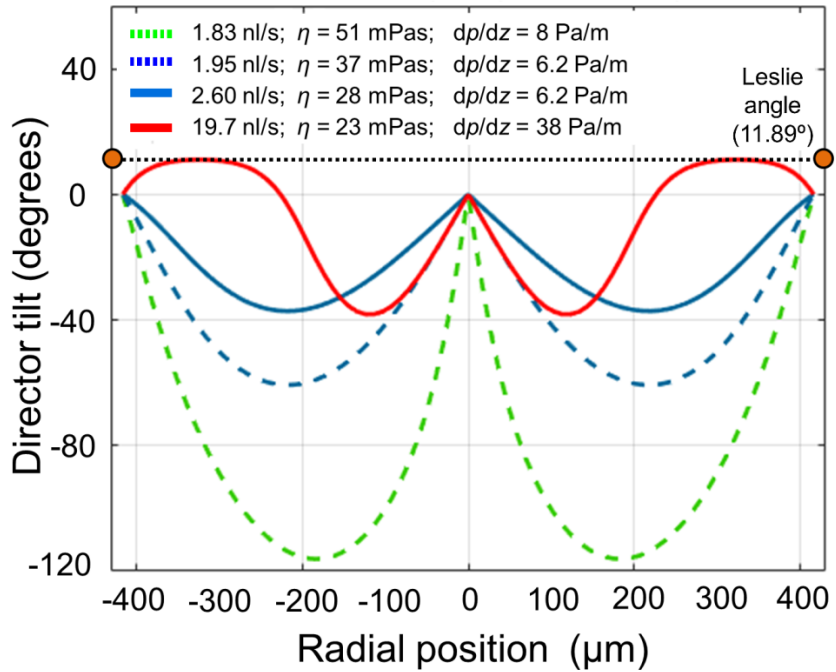

**FIG. 7. Flow-induced director tilt under strong planar anchoring (negative tilt).** Analogous to Fig. 5, the plot presents uniform planar anchoring with negative tilt angles in the vicinity of the capillary center. The maximum tilt angle here is the alignment angle, around 12° which is the same as in Fig. 5. The minimum tilt angle is strongly negative, ~ -120°, leading to a large angular span of the director tilt.

Fig. 9 illustrates how the nature of the bifurcation, i.e. the director topology – manifested as the apparent viscosity and emergent pressure gradients – changes as function of the flow rate and direction (for the homeotropic case). Fig. 9a and b respectively capture the dynamical solutions as a function of the input control parameter, the flow rate. The existence of nematodynamic bifurcation signifies a dynamical transition from one to two solutions. As presented in Fig. 9b, the pressure gradient is plotted as the output, the input parameter being the flow rate. Here one observes a clear transition from one to three solutions. At 2.6 nl/s the pressure is equal to the pressure at the previous flow rate of 1.95 nl/s, plotted using a solid line to indicate that this flow is stable. The pressure gradient increases with increasing flow rates, while the viscosity keeps dropping further. The absolute values of the director tilt angle are smaller than those at smaller flow rates, which is expected due to the decreased apparent viscosity. The largest flow rate depicted in Fig. 7 (19.7 nl/s) interrupts the hitherto increase of the tilt angle slope with increasing flow rates at the capillary center line. Instead, the negative slope recovers again, reaching values comparable to those for the flow rate of 1.95 nl/s. Spatially, the tilt angle reaches its minimum close to the center line, at $r \approx 150$ µm, thereafter increasing sharply to cross the zero tilt angle at $r \approx 226$ µm away from the center. Finally, the director tilt reaches its maximum of ~12° - the Leslie angle - until it drops back to 0° at the capillary boundary. As shown in Fig. 9a, the apparent viscosity drops further, ultimately matching with the apparent viscosity corresponding to the positive tilt angle solution at the same flow rate.

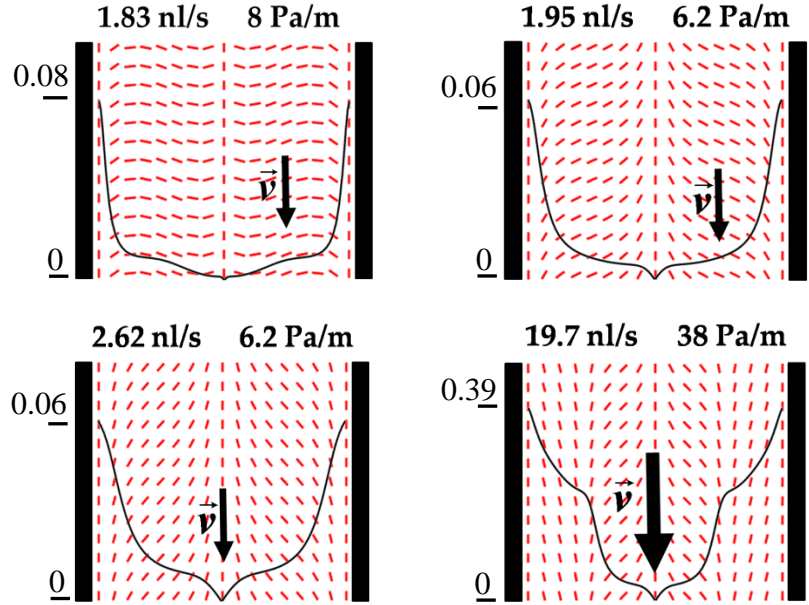

**FIG. 8. Flow-induced director field and emergent shear rates (planar anchoring).** Director field and emergent shear rates for negative the tilt angles in the vicinity of the capillary center. For each of the four plots in Fig. 7, the orientation of the director field is visualized by the orientation of red line-segments. The shear rates are denoted by the black lines overlaying the director field.

The shear rates shown in Fig. 8 can be interpreted as those presented for the Figs. 4 and 6. At regions with zero tilt angles (imposed at the capillary center and at the boundary), one expects the slope of the shear rates to be higher relative to the slopes within rest of the capillary. This is clearly captured in Fig. 8, in contrast to Fig. 4 (lower panels) where the homeotropic anchoring partially causes negative slopes of shear rate near the capillary boundary. At the capillary center, the slope is clearly high for the latter three flow rates. The quicker the tilt angles increase, tighter is the central region with stronger slopes, owing to the increased apparent viscosity due to higher tilt angles. Fig. 9a, furthermore, indicates that the apparent viscosities for the two distinct anchoring states – homeotropic and uniform planar – could potentially converge at high enough flow rates, in agreement with previous experiments [45]. We ascribe the anchoring-specific difference between the apparent viscosities at high *Er* to the anisotropic flow-director coupling close to the capillary boundary. Owing to the no-slip boundary conditions, the boundaries are low Ericksen number regions, thus the follow-director is inherently distinct as one moves from homeotropic to uniform planar boundary conditions. In the limit of strong anchoring conditions, the confinement length scales, relative to the surface extrapolation length, determines the flow-director coupling strength [28]. Thus under finite confinement dimensions, the apparent viscosities for the homeotropic and planar cases could approach each other, however may not merge due to the non-trivial role of the boundary conditions. This could possibly be reversed for large confinement dimensions, where the bulk effects could outweigh the precise effects of surface anchoring. Secondly, for the homeotropic case, one needs to distinguish between the escape configurations of the director field observed. The spontaneous symmetry breaking in the flow-director coupling elicits distinct apparent viscosities corresponding to the positive flow rates (escape direction parallel to flow direction) and the negative flow rates (escape direction antiparallel to flow direction), even for the same *Er* number. At small negative flow rates the viscosity initially goes up with the magnitude of the flow rates, before it starts to decrease. For negative flow rates with strong flow strengths, the director field undergoes considerable elastic deformation. Under uniform planar anchoring, no such flow-director asymmetry exists: consequently, the apparent viscosity remains similar regardless whether the flow started along or opposite to the capillary axis (*z*-axis).

## V. DISCUSSION

### A  Plausibility of the negative tilt angle solutions

For flow-aligning NLC-s like 5CB, the viscous torque tends to align director field toward the Leslie angle in the viscous dominated regime, i.e. at high *Er* [47-49]. However, deviations from the Leslie angle can arise when elastic torque plays a comparable role, for instance at moderate or low Ericksen number flow, as well as in the neighborhood of confining boundaries. Such deviations could also arise at transitions between flow states non-stationary flows, anchoring-induced symmetry breaking or geometric constraints [28,35,41,52]. We do not discard the solutions with negative tilt angles at higher Ericksen numbers (shown in Figs. 3 and 7), instead in section VB, we gather evidence of analogous solutions in experimental micro-flows of NLC-s. The high symmetry of our capillary geometry, combined with the imposed stationarity in our simulations, forbids the possibility of deviations arising due to symmetry breaking of the flow geometry or the existence of non-stationary solutions. Instead, if we consider that our rotationally symmetric solutions exist as coaxial director domains, the intersection between such coaxial domains could serve as new boundary conditions. For instance, let us consider the solution corresponding to the flow rate of 19.7 nl/s (Fig. 7). If we now apply the same pressure gradient of 38 Pa/m, across a cylinder of radius 226 µm, i.e., where the tilt angle of the director crosses the zero line, we arrive at an effective solution for flow through a capillary of 226 µm radius possessing uniform planar anchoring. The solution possesses the same *Er* as the solution with 2.6 nl/s, thus from a LEP perspective, these two solutions are mutually equivalent. The possibility to construct such inner solution domains for smaller Ericksen numbers indicates that the elastic interactions could dominate in the inner part of the flow even at higher Ericksen numbers. Therefore one might find solutions showing negative tilt angles in the inner part at higher Ericksen numbers in experiments with suitably chosen initial conditions. Based on this we hypothesize that the experimental realizations of the alternate flow-induced director orientations (deviated from the Leslie angle) could be enabled or hindered by the lower symmetry of channel flow compared to the cylindrical capillary flows.

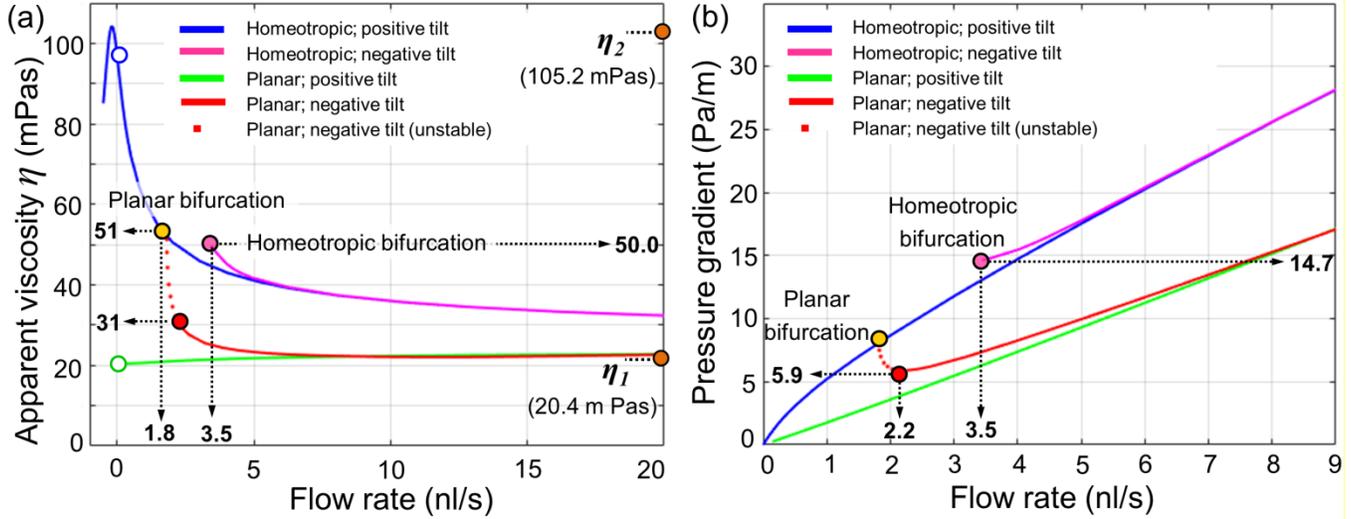

**FIG. 9. Nematodynamic bifurcations in capillary flows.** Apparent viscosity (a) and pressure gradient (b) plotted as a function of the flow rate for the director configurations plotted in the Figs. 3, 5 and 7 and visualized in Figs. 4, 6, 8, respectively for homeotropic and uniform planar anchoring conditions. The values of the Miesowicz viscosities are explained in Section III.

**B  Comparison with experimental results**

Below we discuss two microfluidic experiments in the context of the negative tilt angles suggested in the previous Sections. Fig. 10 shows experimental (left column) and simulated (right column) 5CB director profile that emerge within microfluidic devices possessing homeotropic anchoring [28]. The top left panel shows polarized optical image, while the two bottom images were obtained by confocal fluorescence polarization microscopy. The director configuration is depicted at the bottom right panel by thin white lines. The director orientation exhibits qualitative similarities to the alternate orientation solution presented in Fig. 3 (for the flow rate of 32 nl/s). Close to the channel walls, the director is oriented close to the Leslie angle, while in the vicinity of the channel center, the director bends against the imposed flow field. However the transition region between the positive and the negative tilt angles reported in the experiment differ from our simulated solution: In our numerical solution the director at this transition region is oriented parallel to the flow direction, in the experiment perpendicular.

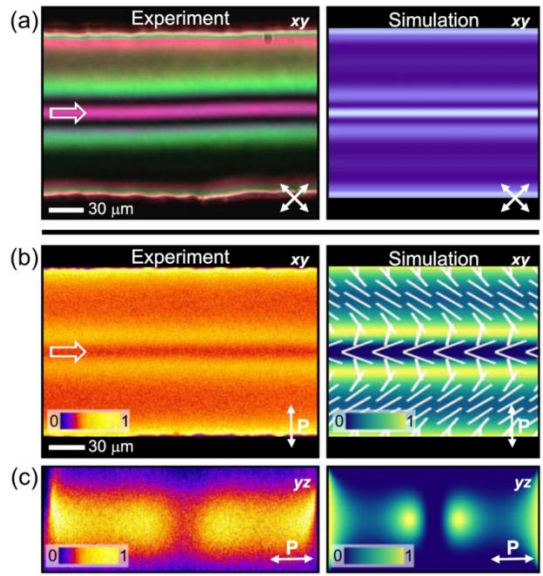

**FIG. 10. Tunable nematic director in microfluidic flows.** Experimental and simulated microfluidic flow and director profiles for nematic 5CB [28]. The flow direction is from left to right, indicated by arrows. The director orientation is indicated by the white lines (right column, middle panel).

In Fig. 11, the results of viscosimetric experiments on 5CB in a channel with a large aspect ratio (2mm x 15µm) and a length of 4 cm are shown [45]. The viscosity is plotted against the pressure difference between the channel inlet and outlet. The red dots indicate the results for planar anchoring which is relevant in our context. For the smallest pressure difference, a viscosity ~25 mPas was observed, which is comparable to the Miesowicz viscosity $\eta_1$, which one expects in the limit of zero flow rate. At a pressure difference of 20 mbar, a clear viscosity maximum of 50 mPas was reported, corresponding to $Er \approx 10$. Upon increasing the pressure difference further, the apparent viscosity falls monotonically, exhibiting expected shear-thinning rheology. While we did not perform corresponding simulations for the channel flow here, we do expect that such simulations would yield similar results, as qualitatively suggested by our simulations for capillary flows (Fig. 9a). Our simulations reproduce a viscosity peak at $Er = 10$ for negative tilt angles, however not for the positive ones. At smaller flow rates, i.e. lower pressure gradients, negative tilt angles are ruled out. Considering the variability in experimental anchoring conditions, we mimicked weak (or variable) anchoring by allowing slightly negative tilt angles at the boundary. This lowered the threshold flow rate for the onset of alternate solution with negative tilt angles. Additionally, we do not exclude that the tilt angles switch sign with changes in the flow rates, underscoring the non-stationarity and perturbations in the flow and director fields. Depending on the manner in which the flow rate is changed, one could induce controlled switching to elicit specific flow-induced solutions. We interpret that the quantitative match observed in the viscosity maximum at $Er \approx 10$ as a promising evidence for the occurrence of negative tilt angles in experiments, however detailed experimental and theoretical investigations will clearly be needed to uncover the underlying nematodynamic details.

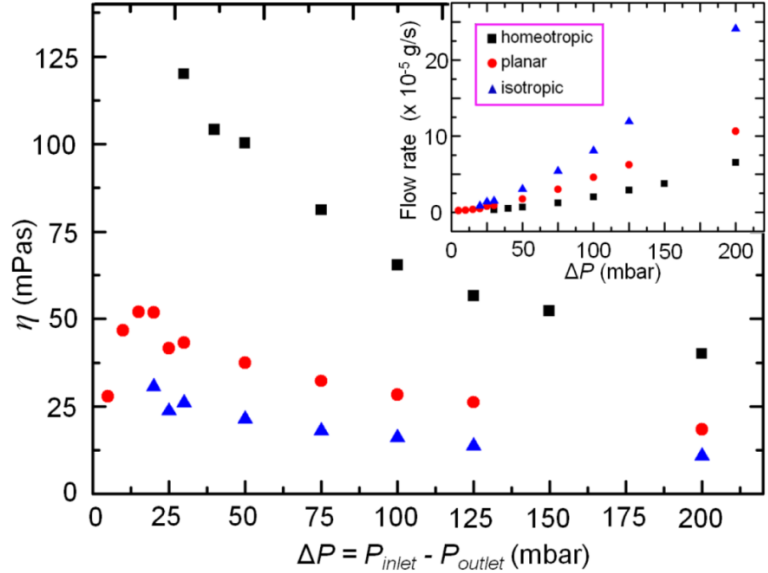

**FIG. 11. Dynamic viscosity of flowing nematic 5CB.** Taken from [45], viscosimetric measurements in a high aspect ratio microchannel filled with 5CB for different anchoring and shear rates. The red dots indicate results obtained for the uniform planar anchoring, as investigated in the current work (Figs. 5-8). A quantitative match has been found for the existence of viscosity maximum at 20 mbar pressure difference.

**VI. Conclusion and Perspectives**

We presented an overview on the stationary flow of aligning rod-shaped nematic liquid crystal flowing within circular capillaries under strong homeotropic and strong uniform planar anchoring. We spanned the elastic and the viscous dominated regimes, i.e. from small to large Ericksen numbers. Our numerical simulations are based on the Leslie-Ericksen-Parodi theory, with relevant modifications and conditions due to the symmetry of a circular capillary. In contrast to the low complexity of the input, the results show a high complexity including non-monotonic viscosity variation, multiple solutions (having both stable and unstable arms) and nematodynamic bifurcations. Our simulations reveal that the flow field bends the director field towards the Leslie angle for both the anchoring types considered here. The higher the Ericksen number, the closer this convergence is. The employed material parameters possess common properties for rod like nematics, in particular its anisotropic viscosity approaches the minimum for a director field parallel to the flow. Consequently, the viscosities are the lowest for planar anchoring at small flow rates, i.e. when the director field is nearly parallel to the flow direction, increasing monotonically for larger flow rates when the Leslie angle is approached throughout the capillary space. For homeotropic anchoring conditions, the opposite phenomenon is observed, with the exception of flows which are antiparallel to the direction of the static escape configuration. The anisotropic viscosity influences the radial variation of the nematic flow, i.e. its shear rate profiles, by deviating them from the usual linear profile of the well-known Hagen-Poisseulle flows. At regions with a small angle between flow and director orientation, the shear rates deviate towards larger values, and at angles close to 90° the shear rates approach smaller values.

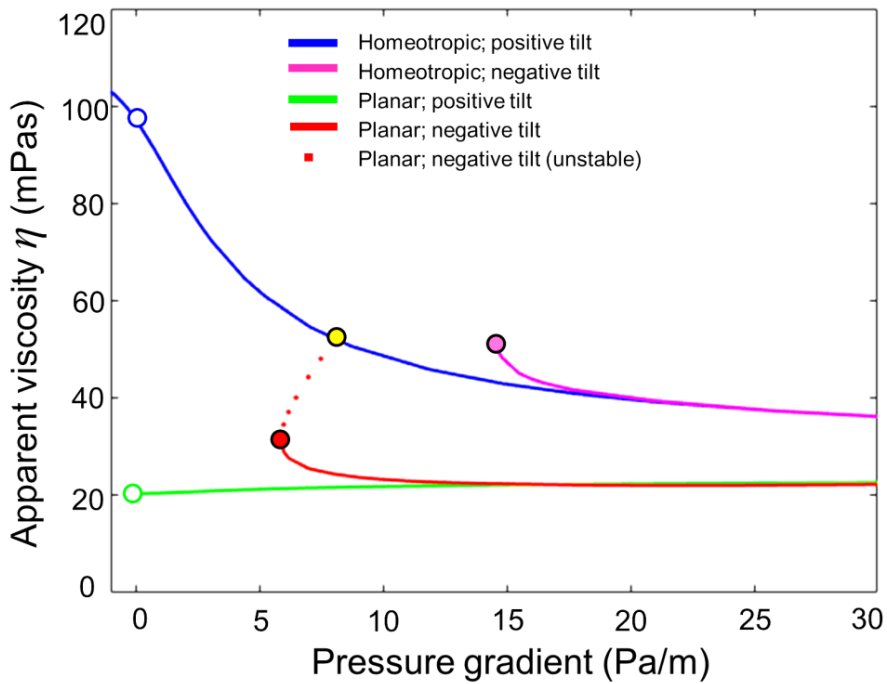

**FIG. 12. Harnessing nematodynamic bifurcations in cylindrical capillaries.** The nonlinear rheological attributes of NLCs could offer a novel handle to control and manipulate microscale flows. Through a careful choice of material and flow parameters, multi-stable solutions in shear rates and director fields could be obtained at will.

The highlight of this paper is the discovery of anchoring-dependent nematodynamic bifurcation, which was captured *in silico* within cylindrical capillaries possessing strong homeotropic and uniform planar anchoring conditions. By spanning two orders of flow rates, we identify anchoring-specific threshold flow rates below which alternate flow-director solution emerges (in addition to the expected director field orientation). The bifurcation coordinates on a pressure-viscosity parameter space depend on the nature of surface anchoring on the walls of the cylindrical capillary, and on the driving mechanism of the flow. For strong homeotropic anchoring conditions, the flow bifurcation was found to occur at a pressure gradient of 14,7 Pa/m (for nematic 5CB), with corresponding flow rate 3.5 nl/s and apparent viscosity 50 mPas. Correspondingly, for Ericksen numbers $Er \geq 11.2$, the second solution emerges. The second solution does not merge with the first solution at the onset threshold, instead its director field and its viscosity are distinct at the onset point. A similar observation was found for discotic liquid crystals before [35]. Close to the capillary center line, the director field orients opposite to the alignment angle. Interestingly, for the planar boundary conditions, the bifurcation coordinates were determined, in addition, by the flow driving mechanism (pressure-driven flows v/s volume-driven flows). For the pressure-driven flows, the bifurcation occurs at 5.9 Pa/m (corresponding flow rate is 2.2 nl/s) and apparent viscosity 31 mPas; whereas for volume-driven flows, the bifurcation takes place at a flow rate of 1.8 nl/s (corresponding pressure gradient is 8 Pa/m) and an apparent viscosity of 51 mPas. In terms of Ericksen numbers, two additional solutions which diverge from each other appear at $Er \geq 5.4$. The solutions do not merge at onset point, with the director fields oriented against the alignment angle. The director field of the additional solutions is bended against the Leslie angle. Overall, the onset of the anchoring-dependent apparent viscosity of nematic 5CB at a given pressure gradient (or flow rate) is coupled to the alternate local director orientations with similar energetic costs, as summarized in Fig. 12. Our numerical results indicate promising agreement with previously reported experiments on nematic microflows within moderate and high aspect ratio microfluidic devices possessing appropriate surface conditions [28,45]. Cylindrical capillary experiments are underway and the results will be reported elsewhere.

The nonlinear rheology of nematic 5CB is summarized in Fig. 12. The multiplicity, the moderate shear thickening for planar anchoring, the strong shear thickening for both anchoring types and particularly the shrinking flow rates for rising pressure gradients could offer novel handle to control and manipulate microscale flows. Furthermore, by expanding the choice of material parameters, alongside confinement geometry and dimensions, one could tailor multi-stable solutions of shear rates at a given Ericksen number. For sufficiently high shear rates, we expect the experimental viscosities for homeotropic anchoring to be lower than our LEP-based prediction (Figs. 9, 12) for two main reasons: Firstly, at high shear rates the surface extrapolation length even of a strong anchoring will come into play and the molecules close to the surface will get differentially tilted toward the flow direction. Secondly, the gradients of the tilt angles at the capillary boundary become larger for higher shear rates, which at a certain magnitude (so far unknown) is expected to lower the local order parameter, inducing a local drop in the nematic elasticity. The first case can enable surface anchoring characterizations macroscopically (without the need for polarization microscopy). The second case can reveal the bulk material properties of the nematics, advancing our insights about the limits of the LEP theory.

**Acknowlegement**

This work was supported by the ATTRACT Investigator Grant (Grant no. A17/MS/ 11572821/MBRACE) and the FNR-CORE Grant (Grant no. C19/MS/13719464/TOPOFLUME/Sengupta) to A.S from the Luxembourg National Research Fund. The authors thank J. Dhar for lively discussions during different phases of this work.